\documentclass[a4paper,12pt]{article}

\usepackage{amsmath,amssymb,amsfonts}% Typical maths resource packages
\usepackage{graphicx}% Packages to allow inclusion of graphics
\usepackage{color}% For creating coloured text and background
\makeatletter
\@addtoreset{equation}{section}
\renewcommand{\theequation}{\thesection.\@arabic\c@equation}
\makeatother
\usepackage{hyperref}% For creating hyperlinks in cross references
\usepackage{cite}
\usepackage{caption}
\definecolor{red}{rgb}{1,0,0}% Standard colours red, green, blue
\definecolor{green}{rgb}{0,1,0}
\definecolor{blue}{rgb}{0,0,1}
\definecolor{darkblue}{rgb}{0,0,0.5}
\definecolor{lightblue}{rgb}{.5,.5,1}
\definecolor{lightgray}{gray}{.87}% How you can define your own greys
\definecolor{Dark}{gray}{.20}
\definecolor{pink}{rgb}{.95,0.82,0.92}% How you can define your own colours
\definecolor{yellow}{rgb}{1,1,0}
\definecolor{lightyellow}{rgb}{1,1,.5}
\definecolor{purple}{rgb}{0.7,0,0.85}
\definecolor{darkgreen}{rgb}{0,0.5,0}
\definecolor{orange}{rgb}{0.8,0.2,0.2}
\def \be {\begin{equation}}
\def \ee {\end{equation}}
\def \bea {\begin{align}}
\def \eea {\end{align}}
\def \nn {\nonumber}

\def \rr {\raise.35ex\hbox{\small $\prime$}\kern-.17em{\mbox{\large $\imath$}}}
\def \del {\partial}
\def \dels {\partial\kern-.5em / \kern.5em}
\def \As {{A\kern-.5em / \kern.5em}}
\def \Ds {D\kern-.7em / \kern.5em}

\def \eps {\epsilon}

\def \lam {\lambda}

\def \th {\theta}

\newcommand{\detail}[1]{}

\newcommand{\hide}[1]{}

\newcommand{\explanation}[1]{}

\pdfoutput=1

\begin{document}

\pagestyle{plain}

\begin{titlepage}

\begin{flushright}
\vspace*{-24pt}
\small\tt
OU-HET 1011
\vspace{24pt}
\end{flushright}

\begin{center}

\noindent
\textbf{\LARGE
Trapping Horizon and Negative Energy \\
%\vskip.6cm
}
\vskip .5in
{\large 
Pei-Ming Ho${}^{a}$%
\footnote[1]{e-mail: pmho@phys.ntu.edu.tw},
%Hikaru~Kawai${}^b$
%\footnote[2]{e-mail: hkawai@gauge.scphys.kyoto-u.ac.jp},
Yoshinori Matsuo${}^{b}$%
\footnote[2]{e-mail: matsuo@het.phys.sci.osaka-u.ac.jp}%,
%Shu-Jung Yang${}^{a}$
%\footnote[3]{e-mail: dodolong0619@gmail.com}
}
\\
\vskip 10mm
{\sl 
${}^{a}$
Department of Physics and Center for Theoretical Physics, \\
National Taiwan University, Taipei 106, Taiwan,
R.O.C. 
\\
\vskip 3mm
${}^{b}$
Department of Physics, Osaka University, \\
Toyonaka, Osaka 560-0043,
Japan
}

\vskip 3mm
\vspace{60pt}
\begin{abstract}

Assuming spherical symmetry
and the semi-classical Einstein equation,
%%%+04/21
%and certain niceness conditions
%on the curvature tensor,
we prove that,
%%%+05/01
for the
%reference frame
observers
%%%-05/01
on top of the trapping horizon,
%YM-4/25
the vacuum energy-momentum tensor
%in the absence of classical matter 
%-4/25
is always that of 
an ingoing negative energy flux at the speed of light
with a universal energy density ${\cal E} \simeq - 1/(2\kappa a^2)$,
%YM-4/25
(where $a$ is the areal radius of the trapping horizon),
%-4/25
%corresponding to a constant rate of change
%${\cal P} \simeq - \frac{2\pi}{\kappa}$
%in the total energy under the apparent horizon,
which is responsible for the decrease in the black hole's mass over time.
This result is independent of the composition of the collapsing matter
and the details of the vacuum energy-momentum tensor.
%%%+05/01
The physics behind the universality of this quantity ${\cal E}$
and its surprisingly large magnitude will be discussed.
%%%-05/01

\end{abstract}
\end{center}

\end{titlepage}

\baselineskip 18pt

%\noindent\rule{\textwidth}{1pt}

%\tableofcontents

%\vskip 12pt

%\noindent\rule{\textwidth}{1pt}

\setcounter{page}{1}
\setcounter{footnote}{0}
\setcounter{section}{0}

\section{Introduction}

The spacetime geometry for the dynamical process
of a gravitational collapse is hard to solve analytically
from the semi-classical Einstein equation
if the back reaction of the vacuum energy-momentum tensor
is included.
The spacetime region far away from the black hole
can be well described by the outgoing Vaidya metric
(including the back reaction of the Hawking radiation),
and the region well inside the collapsing matter
by the ingoing Vaidya metric
(ignoring the vacuum energy-momentum tensor).
But the geometry of a neighborhood of the horizon
has only been studied numerically or 
%YM-4/4
%with various assumptions.
with various approximations. 
%-4/4

For instance, 
for the vacuum energy-momentum tensor 
given by the DFU model \cite{Davies:1976ei},
the numerical analysis was carried out in Ref.\cite{Parentani:1994ij},
and the analytic solution was computed in Ref.\cite{Ho:2018jkm}
with the assumption of weak time-dependence.

In this work,
we zoom into the neighborhood of the trapping horizon
and study the implication of its geometric properties
on the energy-momentum tensor
through the semi-classical Einstein equation,
assuming spherical symmetry for simplicity.
%we define a gauge-invariant geometric quantity
%on the trapping horizon.
There is no assumption about 
the details of the vacuum energy-momentum tensor
%%%+4/21
except the implicit assumption of an $\hbar$-expansion 
%YM-4/25
with the absence of the zero-th order (classical) term 
%-4/25
%but merely some niceness conditions \cite{Mathur:2009hf}
(see eqs.\eqref{Rthth<<1} and \eqref{R2<1} below).
%at the trapping horizon.
%%%-04/21
Surprisingly,
a universal expression (eq.\eqref{T-approx}),
which is uniquely determined by the black-hole mass,
applies to the vacuum energy-momentum tensor
at the trapping horizon.
It corresponds to a light-like ingoing negative energy flux
with the energy density (eq.\eqref{E-})
\begin{align}
T_{\xi\xi}  \simeq - \frac{1}{2\kappa a^2(u)},
\label{Txixi-0}
\end{align}
where $\xi$ is the normalized tangent vector
on the trapping horizon.
%YM-4/25
%%%+04/21

%%%+05/01
Apart from the fact that a universal expression 
of the energy-momentum tensor is derived,
another surprise is that
the expression above \eqref{Txixi-0}
%violates $\hbar$-expansion 
diverges in the limit $\hbar \rightarrow 0$.
%but the first order correction of the expansion becomes 
%comparable to the zero-th order,%
%\footnote{%
%Here, the zero-th order term itself is zero 
%because there are no classical matter in the vacuum.
%}
Correspondingly,
the Einstein tensor
\begin{align}
%%%PM04/10
G_{\xi\xi} 
%\ll 
\sim \mathcal{O}\left(\frac{1}{a^2}\right)
%%%04/10
\end{align}
does not vanish in the limit $\hbar \rightarrow 0$.
(Recall that the Einstein tensor vanishes identically 
for the Schwarzschild metric.)
Rather,
$G_{\xi\xi}$ has a magnitude of the same order 
(but opposite in sign)
as the region occupied by ordinary matter.
We will explain below how this seemingly paradoxical result
is in fact compatible with the $\hbar$-expansion of
the semi-classical Einstein equation.
%which is in general determined by
%the dimension of a gauge-invariant quantity
%and the Schwarzschild radius $a$.
%%%-04/21
%-4/25
%%%-05/01

The same result about the vacuum energy-momentum tensor
was first obtained in a previous work \cite{Ho:2019cfw},
which was restricted to the conventional model of black holes
and only the ingoing negative energy flux was included
in the vacuum energy-momentum tensor as an approximation.
This work is its generalization to a much greater class of models.

In Sec.\ref{SEE},
we solve the semi-classical Einstein equation
\footnote{
By semi-classical Einstein equation,
we mean that the energy-momentum tensor
of the vacuum does not have to vanish,
and that the weak energy condition can be violated.
}
via a coordinate expansion around the trapping horizon.
%%%+04/21
The magnitudes of the quantum corrections to relevant geometric quantities
are estimated
%The niceness conditions that are needed in the calculation
%are spelled out
in Sec.\ref{Nice},
%%%-04/21
which lead to the universal results about 
the vacuum energy-momentum tensor observed 
on the trapping horizon in Sec.\ref{Universal}.
In Sec.\ref{Comments},
we comment on the implications of this result
to the problem of the holographic principle.

\section{Einstein Equations Around Trapping Horizon}
\label{SEE}

Assuming spherical symmetry,
the most general metric in 4D is of the form
\be
ds^2 = - C(u, v) du dv + r^2(u, v) d\Omega^2,
\label{metric}
\ee
where $d\Omega^2 = d\th^2 + \sin^2\th d\phi^2$
is the metric on a 2-sphere.
This metric \eqref{metric} does not uniquely fix
the choice of coordinates.
%YM-4/4
A transformation such as
\begin{align}
u \rightarrow u'(u),
\qquad
v \rightarrow v'(v) , 
\label{gauge-transf}
\end{align}
does not change the form of the metric \eqref{metric},
with $C(u,v)$ transforming as 
\begin{equation}
 C(u,v) \to C'(u',v') = C(u,v) \left(\frac{du'}{du}\right)^{-1} \left(\frac{dv'}{dv}\right)^{-1} \ . 
\end{equation}
This is the residual gauge symmetry %transformation
of %the symmetry of 
diffeomorphism %in general relativity
for the gauge condition in which the metric is of the form \eqref{metric}.
%-4/4

\subsection{Trapping Horizon}

The spacetime with spherical symmetry is foliated
by symmetric 2-spheres labelled by the coordinates $(u, v)$.
A symmetric 2-sphere $S$ is a trapped surface 
if both outgoing and ingoing null geodesics normal to $S$
have negative expansions,
that is,
\begin{align}
\del_v r(u, v) < 0
\quad
\mbox{and}
\quad
\del_u r(u, v) < 0.
\label{delvr<0}
\end{align}
A 4D continuum of trapped surfaces form a trapped region,
and its boundary is a trapping horizon.
A space-like slice of the trapping horizon,
which is a marginally trapped surface,
is called an apparent horizon.

For the space outside the trapped region,
we assume that
\begin{align}
\del_v r(u, v) > 0
\quad
\mbox{and}
\quad
\del_u r(u, v) < 0
\label{delvr>0}
\end{align}
%YM-4/4
(the same as in Minkowski space).
%This holds outside the trapped region both at large distances
%and possibly around the origin
%if the origin is not included in the trapped region.
Here, we assume that the geometry is asymptotically flat, 
or has the same property \eqref{delvr>0} as the Minkowski sapce in the asymptotic region. 
%%%+04/21
%The condition \eqref{delvr>0} holds also around the origin 
%if it is not included in the trapped region.
%-4/4
%
%Notice that,
Since by definition the areal radius cannot be negative,
%the origin $r = 0$ is always a local minimum,
%so
%%%-04/21
in principle eq.\eqref{delvr>0}
is always satisfied at $r = 0$
as long as the trajectory of the origin is time-like.
This means that at a given instant of time,
there are two apparent horizons
as two concentric 2-spheres \cite{Frolov:1981mz,Roman:1983zza,Hayward:2005gi}.

%YM-4/4
%Combining eqs.\eqref{delvr>0} and \eqref{delvr<0},
As eq.\eqref{delvr<0} is satisfied inside the trapping horizon 
while eq.\eqref{delvr>0} holds outside, 
%-4/4
we must have
\begin{align}
\del_v r = 0
\label{delvr=0}
\end{align}
on top of the trapping horizon.

\begin{figure}[h]
\vskip0em
\center
\includegraphics[scale=0.4,bb=0 10 450 500]{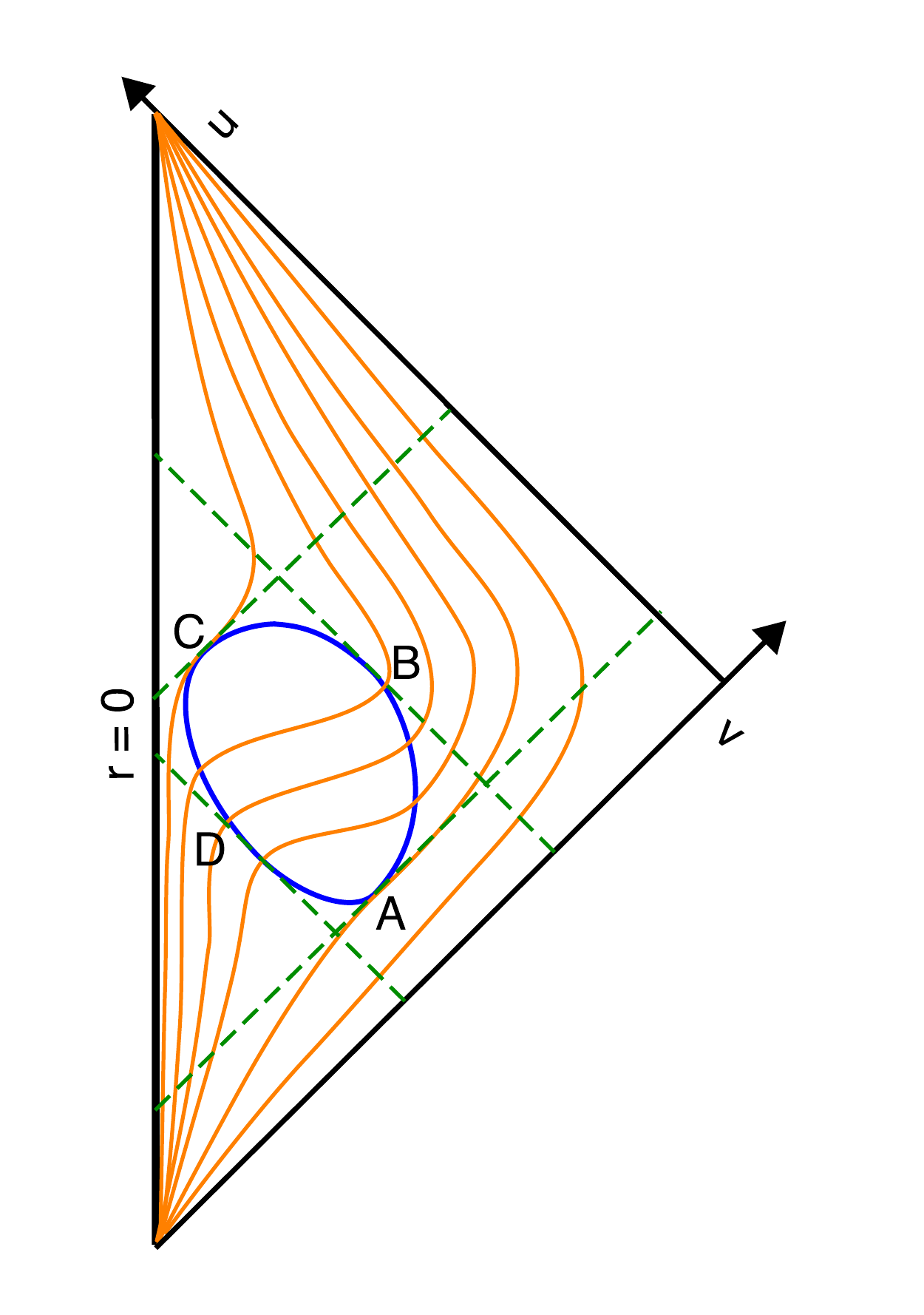}
\caption{\small
The schematic Penrose diagram with a trapping horizon.
\\
The trapping horizon is shown as the solid curve (in blue).
The extremal points on the trapping horizon
with the extrema of the coordinates $u$ and $v$
are marked as $A$, $B$, $C$ and $D$.
The dashed lines (in green) are constant-$u$ and constant-$v$ lines
tangent to the trapping horizon.
The constant-$r$ curves (in orange) are tangent to constant-$u$ lines
on the trapping horizon.
}
\label{trapping-horizon-1}
\vskip0em
\end{figure}

As a finite closed space,
the trapping horizon has
maximal and minimal values of their $u$, $v$ coordinates.
Let us denote by $A$ and $C$ the points
%on the trapping horizon
with the minimal and maximal values of the $u$-coordinate,
respectively.
(See Fig.\ref{trapping-horizon-1}
\cite{Frolov:1981mz,Roman:1983zza,Hayward:2005gi}.)
Similarly,
we denote by $B$ and $D$ the points 
on the trapping horizon with
the maximal and minimal values of the $v$-coordinate.

We will specify an open segment of the trapping horizon as
$AB$, $ABC$, etc.,
where $AB$ refers to the segment on the trapping horizon
between the points $A$ and $B$,
and $ABC$ refers to the segment from $A$ to $C$ through $B$.
The segment $DAB$ is called the outer trapping horizon,
and $BCD$ the inner trapping horizon
in Ref.\cite{Hayward:2005gi}.

Due to eqs.\eqref{delvr<0} and \eqref{delvr>0},
we have
\begin{align}
%ABC: & \quad
%\del_v^2 r > 0,
%\label{dv2r>0}
%\\
%ADC: & \quad
%\del_v^2 r < 0,
%\label{dv2r<0}
%\\
%DAB: & \quad
%\del_u\del_v r < 0,
%\label{dudvr<0}
%\\
%BCD: & \quad
%\del_u\del_v r > 0.
%\\
AB: & 
\quad
\del_v^2 r > 0,
\quad
\del_u\del_v r < 0,
\label{AB}
\\
BC: &
\quad
\del_v^2 r > 0,
\quad
\del_u\del_v r > 0,
\label{BC}
\\
CD: &
\quad
\del_v^2 r < 0,
\quad
\del_u\del_v r > 0,
\label{CD}
\\
DA: & \quad
\del_v^2 r < 0,
\quad
\del_u\del_v r < 0.
\label{DA}
\end{align}
Exceptions are the degenerate cases
in which the 2nd derivatives of $r$ vanishes,
and the inequalities hold for higher derivatives of $r$.
\footnote{
For instance,
the condition
$\del_v^2 r > 0$
on the segment $AB$
can be replaced by
\be
\del_v^2 r = \del_v^3 r = 0, 
\quad \mbox{and} \quad
\del_v^4 r > 0.
\ee
}

\subsection{Expansion Around Trapping Horizon}

The trajectories $ABC$ and $ADC$ of the trapping horizon 
can be specified by their $v$-coordinates
$v_{ABC}(u)$ and $v_{ADC}(u)$,
respectively.
We will use $v_0(u)$ to represent both
%YM-4/4
%$v_{ABC}(u)$ and $v_{ADC}(u)$.
$v_{ABC}(u)$ and $v_{ADC}(u)$.
%-4/4
On a small neighborhood of the segment $ABC$ or $ADC$,
we can expand the metric functions $C(u, v)$ and $r(u, v)$
in powers of $(v-v_0(u))$ as
\begin{align}
C(u, v) &=
C_0(u) + C_1(u)(v-v_0(u)) + \frac{1}{2}C_2(u)(v-v_0(u))^2
+ \frac{1}{6}C_3(u)(v-v_0(u))^3 + \cdots,
\label{C-exp}
\\
r(u, v) &=
a(u) + \frac{1}{2}r_2(u)(v-v_0(u))^2
+ \frac{1}{6}r_3(u)(v-v_0(u))^3
+ \frac{1}{24}r_4(u)(v-v_0(u))^4 + \cdots,
\label{r-exp}
\end{align}
where the linear term in $r(u, v)$ vanishes
because of the trapping horizon condition \eqref{delvr=0}.
According to eq.\eqref{r-exp},
$a(u) = r(u, v_0(u))$ is the areal radius
of the apparent horizon for a given $u$.

For the metric \eqref{metric},
the semi-classical Einstein equations are
\begin{align}
G_{uu} &\equiv
\frac{2\del_u C\del_u r}{Cr} - \frac{2\del_u^2 r}{r}
= \kappa T_{uu},
\label{EEuu}
\\
G_{vv} &\equiv
\frac{2\del_v C\del_v r}{Cr} - \frac{2\del_v^2 r}{r}
= \kappa T_{vv},
\label{EEvv}
\\
G_{uv} &\equiv
\frac{C}{2r^2} + \frac{2\del_u r\del_v r}{r^2} + \frac{2\del_u\del_v r}{r}
= \kappa T_{uv},
\label{EEuv}
\\
G_{\th\th} &\equiv
\frac{2r^2}{C^3}\left(\del_u C\del_v C - C\del_u\del_v C\right)
- \frac{4r}{C}\del_u\del_v r
= \kappa T_{\th\th},
\label{EEthth}
\end{align}
where the coupling constant $\kappa$ is related 
to the Newton constant $G_N$ via $\kappa = 8\pi G_N$.
Due to spherical symmetry,
we have $G_{\phi\phi} = G_{\th\th}\sin^2\th$
and $T_{\phi\phi} = T_{\th\th}\sin^2\th$.
Here the energy-momentum tensor is 
the full tensor including both classical and quantum contributions.

Solving the semi-classical Einstein equations \eqref{EEvv} and \eqref{EEuv}
at the lowest order,
we find
\begin{align}
r_2(u) &=
- \frac{1}{2}\kappa a(u) T^{(0)}_{vv}(u),
\label{r2}
\\
\dot{v}_0(u) &=
\frac{C_0(u) - 2\kappa a^2(u) T^{(0)}_{uv}(u)}{- 2\kappa a^2(u) T^{(0)}_{vv}(u)},
\label{dotv0}
\end{align}
where
\begin{align}
T^{(0)}_{\mu\nu}(u) \equiv T_{\mu\nu}(u, v_0(u)).
\end{align}

On the branch $ABC$,
eqs.\eqref{AB} and \eqref{BC} imply that $r_2 > 0$,
so that the ingoing energy flux must be negative
\begin{align}
T^{(0)}_{vv}(u) < 0
\label{Tvv<0}
\end{align}
according to eq.\eqref{r2}.
On the other hand,
on the branch $ADC$,
eqs.\eqref{CD} and \eqref{DA} imply that
\begin{align}
T^{(0)}_{vv}(u) > 0.
\label{Tvv>0}
\end{align}
It is therefore clear that 
the branch $ABC$ of the trapping horizon 
stays in vacuum
and the other branch $ADC$
typically resides in matter.
At the point $A$ where the two branches join
(see Fig.\ref{trapping-horizon-1}),
we have 
$T^{(0)}_{vv} = 0$.

Furthermore,
on the segment $DAB$,
eqs.\eqref{AB} and \eqref{DA} imply that
$r_2\dot{v}_0 > 0$,
which means that
\be
C_0(u) > 2\kappa a^2(u) T^{(0)}_{uv}(u).
\label{C>Tuv}
\ee
The semi-classical Einstein equation $G_{uv} = \kappa T_{uv}$,
and the spherical symmetry implies that
$G_{uv} = \frac{C}{2r^2}R_{\th\th}$,
so this condition is also equivalent to the geometric condition
\be
R_{\th\th} < 1,
\label{Rthth<1}
\ee
which is typically assumed
%%%+-4/21
in vacuum
%by a niceness condition \eqref{Rthth<<1}
as we will see below
in eq.\eqref{Rthth<<1}.
%%%-04/21

On the other hand,
on the segment $BCD$,
$r_2\dot{v}_0 < 0$, 
so that we need
\be
C_0(u) < 2\kappa a^2(u) T^{(0)}_{uv}(u),
\label{C<Tuv}
\ee
which is equivalent to
\be
R_{\th\th} > 1.
\label{Rthth>1}
\ee

To summarize,
the four segments of the trapping horizon divided by
the extrema of $u$ and $v$ coordinates are 
characterized as follows:
\begin{align}
AB: &\quad T_{vv} < 0, \quad R_{\th\th} < 1,
\\
BC: &\quad T_{vv} < 0, \quad R_{\th\th} > 1,
\\
AD: &\quad T_{vv} > 0, \quad R_{\th\th} < 1,
\\
DC: &\quad T_{vv} > 0, \quad R_{\th\th} > 1.
\end{align}

\begin{figure}[h]
\vskip0em
\center
\includegraphics[scale=0.4,bb=0 0 450 500]{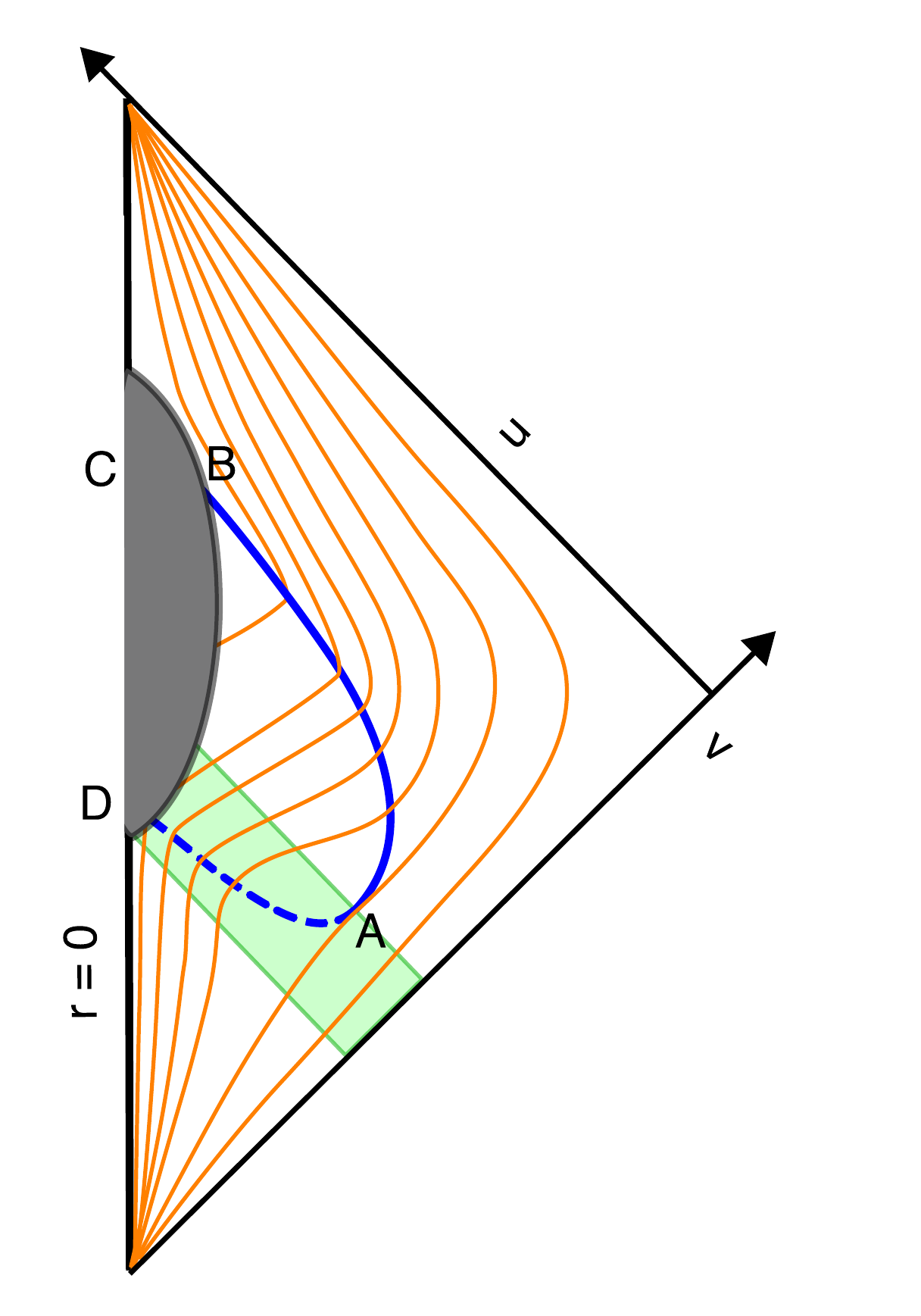}
\caption{\small
The Penrose diagram with a trapping horizon.
\\
The part of the trapping horizon in vacuum is shown
as a solid curve (in blue),
and the part in matter as a dashed curve (in blue),
with the diagonal stripe (in green)
representing the collapsing matter.
The gray area around the origin
represents the high-curvature region behind which
the points $B$, $C$ and $D$ in Fig.\ref{trapping-horizon-1}
are hidden.
The constant-$r$ curves (in orange) are tangent to constant-$u$ lines
on the trapping horizon.
}
\label{trapping-horizon-3}
\vskip0em
\end{figure}

Recall that the Schwarzschild solution has $R_{\th\th} = 0$,
so $|R_{\th\th}| \ll 1$ is expected for an ``uneventful'' horizon in vacuum.
While the segment $CD$ is expected to be completely inside the collapsing matter
where the matter density can be very large to produce
a large spacetime curvature,
%%%+04/21
the exceptionally large $R_{\th\th}$
%%%-04/21
%the violation of some of the so-called ``niceness'' conditions \cite{Mathur:2009hf}
on the segment $BC$
should be limited to the late stage of the evaporation
%YM-4/25
when the size of the black hole is small 
and the Hawking radiation becomes comparable to classical matter,
%-4/25
or when the collapsed matter bounces back.
Hence we modify the Penrose diagram with trapping horizon
in Fig.\ref{trapping-horizon-1}
to Fig.\ref{trapping-horizon-3}.
There,
the segment $BCD$ is hidden in a region 
(the gray area)
where either the collapsing matter has a large energy density around the origin,
or the vacuum energy-momentum tensor is large
at the last stage of the evaporation.

For convenience,
we will refer to the branch of the trapping horizon
represented by the solid curve in Fig.\ref{trapping-horizon-3}
as ``the outer trapping horizon in vacuum'',
and the branch represented by the dashed curve in Fig.\ref{trapping-horizon-3}
as ``the outer trapping horizon in matter''.

\section{%%%+04/21
Magnitude of Quantum Corrections
%Niceness Conditions
%%%-04/21
}
\label{Nice}

Without specifying the details of 
the vacuum energy-momentum tensor,
it is often assumed that,
for a large black hole,
the horizon does not look too different from
the Minkowski vacuum.
For example,
for a black hole with the Schwarzschild radius $a$,
one expects the scalar curvature
%YM-4/25
$R \ll \mathcal{O}(1/a^2)$.
%-4/25
%%%+04/21
%Similarly, 
%other gauge-invariant quantities of (energy) dimension $n$
%should be of order $\mathcal{O}(1/a^n)$ or less.
%Such assumptions are called 
%niceness conditions \cite{Mathur:2009hf}.
%%%-04/21
Throughout this paper,
we shall assume that
\be
\kappa \ll a^2.
\ee

%%%+04/21
%On the other hand,
%it was pointed out in Ref.\cite{Mathur:2009hf} that
%some niceness conditions must be broken if
%the Hawking radiation takes away the information
%of the collapsing matter.
%This was further demonstrated in
%the firewall scenario \cite{firewall}.

In vacuum, 
the energy-momentum tensor involves a factor of $\hbar$,
and the factor $\hbar$ appears
in the semi-classical Einstein equation $G_{\mu\nu} = \kappa T_{\mu\nu}$
only through its product with the Newton constant $\kappa$.
We can thus adopt the convention that $\hbar = 1$
and count the order of $\hbar$ in the $\hbar$-expansion
in terms of the order of $\kappa$ in the $\kappa$-expansion.

For the Schwarzschild metric,
an invariant geometric quantity of dimension $n$ is expected 
to be of order $\mathcal{O}(1/a^n)$ as
the Schwarzschild radius $a$ is the only quantity with nontrivial dimension,
and its quantum correction should be of order
$\mathcal{O}(\kappa/a^{n+2})$ or less.
For geometric quantities that are exactly $0$ for the Schwarzschild metric
(e.g. the scalar curvature),
we can estimate their magnitudes by their quantum corrections.

%For spherically symmetric configurations,
%in terms of the coordinates $(u, v, \th, \phi)$ used in eq.\eqref{metric},
%the niceness conditions demand that
%a geometrical quantity in vacuum of dimension $n$
%invariant under the residual gauge symmetry \eqref{gauge-transf}
%must be of order $\mathcal{O}(1/a^n)$ or less.
%%%-04/21

For example,
on the trapping horizon in vacuum
(the solid blue curve in Fig.\ref{trapping-horizon-3}),
\begin{align}
|R_{\th\th}| &\lesssim \mathcal{O}\left(\frac{\kappa}{a^2}\right)
\ll 1,
\label{Rthth<<1}
\\
|g^{uv} R_{uv}| &\lesssim \mathcal{O}\left(\frac{\kappa}{a^4}\right) 
\ll \mathcal{O}\left(\frac{1}{a^2}\right),
\\
|\det R^{(2)}| &\lesssim \mathcal{O}\left(\frac{\kappa}{a^6}\right) 
\ll \mathcal{O}\left(\frac{1}{a^4}\right),
\label{R2<1}
\end{align}
where $R^{(2)}$ is the 2D-matrix of the Ricci tensor
\be
R^{(2)} \equiv \left(
\begin{array}{cc}
R^u{}_u & R^u{}_v \\
R^v{}_u & R^v{}_v
\end{array}
\right).
\ee
Since $G_{\mu\nu} \equiv R_{\mu\nu} - \frac{1}{2}g_{\mu\nu}R$,
and the Ricci scalar is
$R \equiv 2g^{uv}R_{uv} + 2g^{\th\th}R_{\th\th}$
due to the spherical symmetry,
the conditions above are equivalent to
\begin{align}
\left|
2\kappa a^2(u) T^{(0)}_{uv}(u)
\right|
&\ll C(u, v_0(u)),
\label{Tuv-estimate}
\\
\left|
\kappa T^{(0)}_{\th\th}(u)
\right|
&\ll 1,
\label{Tthth-estimate}
\\
\left|
4\kappa^2 a^4(u) T^{(0)}_{uu}(u, v_0(u))T^{(0)}_{vv}(u)
\right|
&\ll
C^{2}(u, v_0(u)).
\label{TuuTvv-estimate}
\end{align}
These inequalities are invariant under the gauge transformation
\eqref{gauge-transf}.

Eqs.\eqref{Rthth<<1} and \eqref{R2<1}
(or equivalently,
eqs.\eqref{Tuv-estimate} and \eqref{TuuTvv-estimate})
are the only inequalities assumed in this work.
In the conventional model of black holes,
$T_{uu}$ and $T_{uv}$ are of order
$\mathcal{O}(\kappa^2/a^8)$ and $\mathcal{O}(\kappa/a^4)$,
respectively,
at the horizon,
and $|T_{vv}|$ is of order $\mathcal{O}(1/a^4)$,
so these assumptions are satisfied.
But we emphasize that our assumptions are weaker
than the conventional model.
For instance, 
we do not need to assume that $|T_{vv}|$ is much larger than $|T_{uu}|$.

%%%PM04/10
\section{Energy-Momentum Tensor on Trapping Horizon}
\label{Universal}
%%%+04/21
%\subsection{A Universal Result}
%%%-04/21
%%%04/10

In the following,
we focus on the spacetime geometry
in a small neighborhood of 
the outer trapping horizon,
%where the semi-classical Einstein equation is valid,
i.e.
the segment $DAB$ in Fig.\ref{trapping-horizon-3}.
%Hence $v_0(u)$ below only refers to $v_{ABC}$.

Define a time-like vector on the trapping horizon by
\begin{align}
\xi(u) = (\xi^u(u), \xi^v(u))
= \lam(u) (1, |\dot{v}_0(u)|),
\end{align}
where
\be
\lam(u) = \frac{1}{\sqrt{C_0|\dot{v}_0|}}
%= \left[
%\frac{- 2\kappa a^2(u) T^{(0)}_{vv}(u)}
%{C_0(u)(C_0(u) - 2\kappa a^2(u) T^{(0)}_{uv}(u))}
%\right]^{1/2}
\label{lam}
\ee
is chosen such that $\xi(u)$ is of unit length.
The normalized vector orthogonal to $\xi(u)$ in the $u-v$ plane is then
\be
\zeta(u) = (\zeta^u(u), \zeta^v(u)) = \lam(u) (1, -|\dot{v}_0(u)|).
\ee
Together,
$\xi(u)$ and $\zeta(u)$ compose an orthonormal basis of vectors
on the trapping horizon.
On the trapping horizon in vacuum,
$\xi(u)$ is tangent to the trapping horizon.
On the trapping horizon in matter,
$\zeta(u)$ is tangent to the trapping horizon.
%If $\dot{v}_0 < 0$,
%$\xi(u)$ is space-like and $\zeta(u)$ is time-like.

The energy density, energy flow and pressure
for an observer on the trapping horizon
%in this reference frame
are
\begin{align}
T_{\xi\xi} &\equiv T_{\mu\nu}^{(0)}\xi^{\mu}\xi^{\nu}
= \lam^2\left[
T^{(0)}_{uu} + 2|\dot{v}_0| T^{(0)}_{uv} + \dot{v}_0^2 T^{(0)}_{vv}
\right]
\nn \\
&=
\eps
\frac{
(C_0 - 2\kappa a^2 T^{(0)}_{uv})\left((C_0 - 2(1-2\eps)\kappa a^2 T^{(0)}_{uv}\right)
+ 4\kappa^2 a^4 T^{(0)}_{uu}T^{(0)}_{vv}
}
{2\kappa a^2 C_0(C_0 - 2\kappa a^2 T^{(0)}_{uv})},
\label{Txixi}
\\
T_{\xi\zeta} &\equiv T_{\mu\nu}^{(0)}\xi^{\mu}\zeta^{\nu}
= \lam^2\left[
T^{(0)}_{uu} - \dot{v}_0^2 T^{(0)}_{vv}
\right]
\nn \\
&= 
-\eps
\frac{
(C_0 - 2\kappa a^2 T^{(0)}_{uv})^2 - 4\kappa^2 a^4 T^{(0)}_{uu}T^{(0)}_{vv}
}
{2\kappa a^2 C_0(C_0 - 2\kappa a^2 T^{(0)}_{uv})},
\label{Txizeta}
\\
T_{\zeta\zeta} &\equiv T_{\mu\nu}^{(0)}\zeta^{\mu}\zeta^{\nu}
= \lam^2\left[
T^{(0)}_{uu} - 2|\dot{v}_0| T^{(0)}_{uv} + \dot{v}_0^2 T^{(0)}_{vv}
\right]
\nn \\
&=
\eps
\frac{
(C_0 - 2\kappa a^2 T^{(0)}_{uv})\left(C_0 - 2(1+2\eps)\kappa a^2 T^{(0)}_{uv}\right)
+ 4\kappa^2 a^4 T^{(0)}_{uu}T^{(0)}_{vv}
}
{2\kappa a^2 C_0(C_0 - 2\kappa a^2 T^{(0)}_{uv})},
\label{Tzetazeta}
\end{align}
where $\eps$ is the sign of $T_{vv}^{(0)}$,
\be
\eps \equiv \frac{T_{vv}^{(0)}}{|T_{vv}^{(0)}|}.
\ee

%Note that $T_{\xi\zeta}$ is always positive,
%meaning that there is either outgoing positive energy flux
%or ingoing negative energy flux at the apparent horizon.

%%%+04/21
\subsection{A Universal Result}
%%%-04/21

According to the estimate \eqref{Tuv-estimate},
the value of $\dot{v}_0(u)$ \eqref{dotv0} and 
%$\lam$ \eqref{lam},
the energy-momentum tensor \eqref{Txixi}--\eqref{Tzetazeta}
can be simplified as
\begin{align}
\dot{v}_0(u) \simeq
- \frac{C_0(u)}{2\kappa a^2(u) T^{(0)}_{vv}(u)},
\label{dotv0-approx}
\\
%\lam(u) \simeq
%\frac{1}{\sqrt{2\kappa a^2(u)|T^{(0)}_{vv}(u)|}}
T_{\xi\xi} \simeq - T_{\xi\zeta} \simeq T_{\zeta\zeta}
\simeq \frac{\eps}{2\kappa a^2}
\label{T-approx}
\end{align}
in the neighborhood of the trapping horizon.

The main result of this work is the following.
First,
for $T_{vv} < 0$ in vacuum, 
we have
$\dot{v}_0 > 0$ according to eq.\eqref{dotv0-approx},
and so the outer trapping horizon in vacuum is time-like.
More importantly,
the vacuum energy-momentum tensor \eqref{T-approx} for observers 
staying on top of the outer trapping horizon in vacuum
is thus a light-like negative ingoing energy 
with the energy density 
\be
{\cal E}_{-}(u) \equiv T_{\xi\xi} \simeq - \frac{1}{2\kappa a^2(u)},
\label{E-}
\ee
which amounts to the power of
\begin{align}
{\cal P}_{-} \simeq - \frac{2\pi}{\kappa}
\label{P-}
\end{align}
through the outer trapping horizon in vacuum.

Notice that,
due to the spherical symmetry,
${\cal E}_{-}$ is a gauge-invariant quantity,
and it is independent of the details of the quantum field theory
responsible for the vacuum energy-momentum tensor
or the initial state.
It only depends on the assumption of spherical symmetry,
the semi-classical Einstein equation,
and the existence of the trapping horizon.

From the viewpoint of observers on the outer trapping horizon in vacuum,
the mass of the black hole decreases 
not because of the outgoing Hawking radiation,
but because of the ingoing negative energy flow ${\cal P}_{-}$.

%%%+04/21
Notice that the magnitude of the energy density \eqref{E-}
is exceptionally large for vacuum.
How is it possible to have such a large quantum effect?
Let us examine this issue more carefully.

For the classical Schwarzschild metric,
the vacuum energy-momentum tensor $T_{\mu\nu}$ vanishes identically,
hence one would expect that
%%%+05/01
%the energy density vanishes for all observers.
its quantum correction is of order $\mathcal{O}(\hbar)$.
There is however a technical subtlety.
The energy density ${\cal E_-}$ is defined for observers
sitting on top of the horizon.
For the classical black hole,
the horizon is light-like,
so these observers are moving outward at the speed of light.
As a result, 
%regardless of how small $T_{vv}$ is,
%as long as it is non-zero,
%the energy density ${\cal E_-}$ diverges due to the
there is an
infinite Lorentz factor,
%When $T_{vv}$ is exactly $0$,
so that
the energy density ${\cal E_-}$ is strictly speaking ill-defined,
as a product of $0$ and infinity.

If we compute the vacuum expectation value of
the energy-momentum tensor for a certain quantum field
in the Schwarzschild background,
typically $T_{vv}$ is non-zero,
and the energy density ${\cal E_-}$ diverges at the horizon
due to the infinite Lorentz factor.
%%%-05/01
%
\footnote{%
It should be noted that the divergence does not appear 
on the future horizon but on the past horizon 
and at the intersection of two horizons. 
}

Our calculation above shows that,
when the back reaction of the vacuum energy-momentum tensor 
is taken into consideration
so that the Schwarzschild metric is modified,
the energy density ${\cal E}_-$ is ``regularized''
from infinity to a finite value given by eq.\eqref{E-}.
It is therefore not unreasonable to see that
it diverges in the limit $\kappa \rightarrow 0$.

Local gauge-invariants such as $T^{\mu}_{\mu}$ are still small,
vanishing in the classical limit $\hbar \rightarrow 0$.
The energy density ${\cal E}_-$ is a {\em non-local} invariant.%
\footnote{
${\cal E}_{-}$ is not a {\em local} gauge-invariant
as it demands the information about the trapping horizon,
which is uniquely defined only when there is spherical symmetry,
and the spherical symmetry requires nonlocal information
about the spacetime geometry.
}
However, 
recall that Hawking radiation is also
described by a nonlocal term in the effective action.
%%%-04/21

%%%%PM04/10
%{\color{red}
%I am no longer sure about this because
%the time used to define ${\cal P}_{-}$
%is not the same as that used to define $M$.
%We can just remove this paragraph.
%}
%%%%04/10
%As the power ${\cal P}_{-}$ is independent of $a$,
%the time it takes for the observer on the apparent horizon
%to see the complete evaporation of the black hole,
%or rather,
%to see the ingoing negative energy canceling
%the mass of the collapsing matter,
%is always
%\be
%T = \frac{M}{|{\cal P}|} 
%\simeq \frac{\kappa M}{2\pi} 
%= 2a.
%\ee

Similarly,
since $T_{vv} > 0$ inside the collapsing matter,
we have
$\dot{v}_0 < 0$ according to eq.\eqref{dotv0-approx},
and so the outer trapping horizon in matter is space-like.
The matter energy-momentum tensor defined
in the (gauge-invariant) reference frame of ($\xi$, $\zeta$)
is a light-like positive ingoing energy
with the energy density
\be
{\cal E}_{+}(u) \simeq \frac{1}{2\kappa a^2(u)},
\ee
which leads to the power of
\begin{align}
{\cal P}_{+} \simeq \frac{2\pi}{\kappa}
\label{P+}
\end{align}
through the outer trapping horizon in matter.

%%%PM04/10
\subsection{Connection to Surface Gravity}

The reader may find it surprising that
the vacuum energy density observed 
on the outer trapping horizon in vacuum
is given by a universal formula \eqref{E-}
independent of how we define
the quantum field theory responsible for the vacuum energy.
The same result holds for any given energy-momentum tensor $T_{\mu\nu}$
as long as
%%%+04/21
the trapping horizon exists.
%the niceness conditions hold and $T_{vv} < 0$.
%%%-04/21
Such a robustness of the energy density on the horizon is possible
for the following reason.
If we imagine a modification to the energy-momentum tensor
due to a change in the quantum field theory
or a change in the initial state,
the slope $dv/du$ \eqref{dotv0-approx} would change accordingly,
so that the trajectory of the trapping horizon is also changed.
As $T_{\xi\xi}$ is defined with respect to
the tangent vector $\xi$ of the trapping horizon,
it is possible that the changes in $\xi$ and $T_{\mu\nu}$
cancel to keep the value of $T_{\xi\xi}$ invariant.

The comment above explains how it is possible for $T_{\xi\xi}$
to keep a constant value independent of all sorts of variations 
of the underlying theory or states,
but it does not explain why it is given by the specific value \eqref{E-}.
To address the latter issue,
we comment on the relation between
$T_{\xi\xi}$ and the surface gravity. 

%As the energy-momentum tensor at the trapping horizon is universal, 
%it would be related to a geometric properties of the trapping horizon. 
Using the Raychaudhuri equation, 
we have
\begin{align}
 \kappa T_{\xi\xi} 
 \simeq 
 R_{\xi\xi} 
 = 
 - (D^\mu \xi^\nu) (D_\nu \xi_\mu) 
 - \xi^\mu D_\mu (D_\nu \xi^\nu) 
 + D_\nu (\xi^\mu D_\mu \xi^\nu) \ .
\label{Ray}
\end{align}
The second term on the right hand side is 
the time evolution of the expansion $D_\nu \xi^\nu$ along the trapping horizon, 
and the third term would vanish if $\xi$ is the tangent vector of a geodesic
(although $\xi$ is not a geodesic here). 
Notice that,
if $\xi$ is the Killing vector 
at the Killing horizon of a static black hole,
both the second and third terms would vanish,
and the first term is related to the surface gravity. 
Recall that
the surface gravity $\kappa_H$ for a static black hole
is expressed in terms of the Killing vector $\xi$ as 
\begin{equation}
 \kappa_H^2 
 = - \frac{1}{2} (D^\mu \xi^\nu) (D_\mu \xi_\nu) 
 = \frac{1}{2} (D^\mu \xi^\nu) (D_\nu \xi_\mu) \ ,
\end{equation}
which is proportional to the first term on the right hand side of eq.\eqref{Ray}.
%YM-4/15
For the Schwarzschild solution, 
the apparent horizon is identical to the event horizon and 
the surface gravity is given by $\kappa_H = 1/2a$. 
Thus the right hand side of \eqref{Ray} gives the same universal result 
for the Schwarzschild solution as $\xi$ is replaced by 
the generator of the Killing horizon. 

%(It is $\kappa_H = 1/2a$ for the Schwarzschild solution.)
%If the time-dependence of the dynamical process
%does not significantly modify each term in the expression of $T_{\xi\xi}$ above,
%we expect that $T_{\xi\xi}$ at the trapping horizon
%can be estimated as $-2/\kappa$ times the surface gravity squared. 

That is,
the universal result comes from 
the geometric properties of the trapping horizon and its tangent vector, 
which can be related to the surface gravity of the event horizon. 
We are led to a generalization of the surface gravity to 
the dynamical cases without Killing vectors to
\begin{equation}
 \kappa_A^2 \equiv - \frac{1}{2} \kappa \mathcal E_- \ , 
\end{equation}
although it is quite different from the generalizations of 
the surface gravity proposed in the literatures. 
In order to see if the definition above is appropriate, 
the relation to the surface gravity in more general geometries should be examined. 
This is left for future studies. 
%in general,
%even in the absence of spherical symmetry,
%the energy density for observers on top of the trapping horizon in vacuum
%can always be estimated by
%\be
%{\cal E}_- \simeq \frac{-2\kappa_H^2}{\kappa},
%\ee
%where $\kappa_H$ is the surface gravity for the corresponding static solution.
%\footnote{
%The surface gravity for the Kerr-Newman solution of
%mass $M$, charge $Q$ and angular momentum $J$ is
%\be
%\kappa_H = \frac{\sqrt{M^2-Q^2-J^2/M^2}}{2M^2-Q^2+2M\sqrt{M^2-Q^2-J^2/M^2}}.
%\ee
%}
%Similar discussion can be applied to the unit normal vector $\zeta$. 
%%%04/10
%-4/15

\section{Comments}
\label{Comments}

%%%+04/21
%As this work generalizes the results of Ref.\cite{Ho:2019cfw}
%by relaxing the assumptions,
%the physical implications of these results are essentially the same.
%We briefly recapitulate the points relevant 
%to the information loss paradox and the holographic principle.
%%%-04/21

%YM-4/4
From the viewpoint of the observers on top of the trapping horizon in vacuum,
%-4/4
%%%+04/21
We have shown that
%the niceness conditions imply that 
%%%-04/21
the energy-momentum tensor
is always that of an ingoing negative energy flux at the speed of light.
The negative vacuum energy accumulates behind the trapping horizon,
so that the total energy decreases over time.
As the black hole's mass decreases,
the size of the ``neck'' $a(u)$ shrinks,
while the interior space under the neck can still remain large.
The geometry is reminiscent of ``Wheeler's bag of gold''.
The collapsed matter
(together with its information)
stays in the bag.

The low-energy effective theory cannot reliably predict
whether the size of the neck shrinks to $0$,
and whether the bag detaches then.
As long as the neck does not shrink to exactly $0$,
the information inside the bag
is still accessible to outside observers.
However,
from the viewpoint of distant observers,
a black hole with a tiny neck is a remnant
of tiny energy but huge entropy.
Feynman diagrams involving these remnants
are expected to contribute to scattering amplitudes
due to the huge volume of phase space to be integrated over.

Furthermore,
since macroscopic negative energy is allowed to accumulate,
no one should expect the holographic principle to hold.
%This is in agreement with the firewall proposal \cite{firewall}.
%%%+05/01
For reviews on the information paradox,
see e.g. Refs.\cite{Mathur:2009hf} and \cite{Marolf:2017jkr}.
%%%-05/01

If the black-hole remnants are not acceptable,
or if the holographic principle must hold,
%YM-4/24
%YM-4/4
some of the assumptions we made in this paper must be invalid.
%-4/4
%%%PM04/10
%%%+05/01
In this paper, we assumed the presence of the time-like apparent horizon. 
We also assumed that gauge-invariant combinations of 
the vacuum energy-momentum tensor is much smaller than that of classical matter 
such that the geometry is not modified by the quantum correction 
at the leading order of the perturbative expansion, namely, \eqref{Rthth<<1}-\eqref{R2<1}. 
However, the ingoing negative energy which is necessary for the time-like apparent horizon 
is as large as the energy of classical matter (but opposite in sign)
from the viewpoint of an observer on the trapping horizon.
It seems unavoidable to have some gauge-invariant quantities large in vacuum.
%and the perturbative expansion is invalid. 

As we found that the quantum effect in the energy-momentum tensor 
becomes comparable to that of classical matter, 
the assumption that the quantum corrections must be very small 
in the gauge-invariant combinations \eqref{Rthth<<1}-\eqref{R2<1} is
no longer
%very reasonable. 
incontestable.
For the sake of preserving the holographic principle,
%One might demand that
it might be preferable to impose the condition
\be
G_{\mu\nu}\xi^{\mu}\xi^{\nu} = \kappa T_{\xi\xi} \ll 
%%PM04/10
\mathcal{O}\left(\frac{1}{a^2}\right) \ , 
\label{AnotherNiceness}
%%04/10
\ee
%by relaxing some 
instead of all of the conditions \eqref{Rthth<<1}-\eqref{R2<1}. 
Whether the holographic principle can hold
relies on the properties of the vacuum energy-momentum tensor.
%%%-05/01
%It depends on the properties
%of the vacuum energy-momentum tensor if the holographic principle can hold.
%Certain characteristic features of the energy-momentum tensor
%determine whether the holographic principle can hold.
For example,
if it satisfies the null energy condition,
there can be no trapping horizon
%(so the 3rd assumption is violated)
because eq.\eqref{Tvv<0} cannot be satisfied.
%%%+04/21
The KMY model \cite{Kawai:2013mda}
(see also Refs.\cite{Kawai:2014afa}--\cite{Kawai:2017txu})
is such a scenario.

%Alternatively,
%it is possible that there are effective action terms
%(e.g. those related to ${\cal E_-}$)
%that significantly modifies the semi-classical Einstein equation
%around the trapping horizon.
%%%-4/21
%This is the case of the KMY model
%\cite{Kawai:2013mda}--\cite{Kawai:2017txu}.
%There is so far no explicit models of 
%vacuum energy-momentum tensor
%that admits a trapping horizon 
%but a dramatically modified geometry.
%-4/24

%The simplest way to avoid the breakdown of the holographic principle
%is thus to assume that the vacuum energy-momentum tensor 
%satisfies the weak energy condition at least at macroscopic scales.
%It will be interesting to see how
%the vacuum energy-momentum tensor of a certain unified quantum field theory
%can satisfy this criterium.

%A black hole is observationally characterized by 
%a huge red-shift factor near the Schwarzschild radius.
Recall that,
for the Schwarzschild solution,
%there is an infinite blue-shift factor at the event horizon.
%Yet
%YM-4/4
the event horizon consists of 
the future horizon
(i.e. the black-hole horizon)
and the past horizon
(i.e. the white-hole horizon).
%-4/4
It is conventionally assumed that 
only the future horizon is relevant 
to the dynamical process of black-hole
formation and evaporation.
However, 
in the KMY model \cite{Kawai:2013mda}
and the model proposed in Ref.\cite{Stephens:1993an},
the white-hole horizon
plays (at least conceptually) a crucial role.
%%%+04/21
We apply the same calculation to these models in the appendix.
%we show that this class of models always breaks
%the niceness condition \eqref{Rthth<<1}.
%This is expected because there does not have to be information loss
%in those cases.
%%%-04/21

%%%+04/21
%It will
%be interesting to investigate 
%how different types of vacuum energy-momentum tensors
%lead to violations of different niceness conditions,
%and how the violation of different niceness conditions
%may
%lead to different models of black holes
%(e.g. the fuzzballs \cite{FuzzBall}, 
%the KMY model \cite{Kawai:2013mda},
%and the anti-podal identification \cite{Hooft:2018gtw} etc.).

To summarize,
different classes of vacuum energy-momentum tensor
(corresponding to different classes of quantum field theories)
lead to different models of black holes.
Some of the models are compatible with the holographic principle,
e.g. the fuzzballs \cite{FuzzBall}, 
the KMY model \cite{Kawai:2013mda},
the firewall \cite{firewall},
and the anti-podal identification \cite{Hooft:2018gtw},
and other models are not compatible with the holographic principle
(e.g. the conventional model).
Yet regardless of whether you believe in the holographic principle,
we emphasize the crucial role played by the quantum effect
of the vacuum energy-momentum tensor.
%%%-04/21
%%%04/10

\section*{Acknowledgement}

The authors would like to thank 
Chong-Sun Chu,
Hikaru Kawai,
Samir Mathur,
Yutaka Matsuo,
Wen-Yu Wen,
and Yuki Yokokura
for discussions.
The work is supported in part by
the Ministry of Science and Technology, R.O.C.
(project no. 107-2119-M-002 -031 -MY3)
and by National Taiwan University
(project no. 105R8700-2).
The work of Y.M. 
is supported in part by 
JSPS KAKENHI
Grants No.~JP17H06462.

\appendix

\section{Appendix: An Alternative Scenario}

The dynamical analogue of the white-hole horizon
is an anti-trapping horizon.
At large distances where the curvature is small,
we have eq.\eqref{delvr>0}.
We would have an anti-trapped region if
\be
\del_u r > 0,
\quad \mbox{and} \quad
\del_v r > 0
\ee
in part of the spacetime.
The boundary of the anti-trapped region is 
then an anti-trapping horizon.

Here we consider alternative scenarios
\cite{Stephens:1993an,Kawai:2013mda}
in which the anti-trapping horizon 
plays an important conceptual role in the following sense
although there is actually no anti-trapped region.
There would be an anti-trapping horizon
in the region occupied by the collapsing matter
if the geometry is given by the analytic continuation
of the geometry in vacuum outside the collapsing matter.

%%%
%For Minkowski space and Schwarzschild metric,
%$\del_u r$ is negative and $\del_v r$ is positive,
%as $u$ is the outgoing light-cone coordinate
%and $v$ the ingoing light-cone coordinate.
%If there is a region bounded by a certain trajectory $u = u_0(v)$
%where
%\be
%\del_u r > 0
%\qquad \mbox{and} \qquad
%\del_v r > 0,
%\ee
%the region is {\em anti-trapped},
%and $u = u_0(v)$ appears similar to the white-hole horizon (past horizon).

\begin{figure}
\vskip-2em
\center
\includegraphics[scale=0.3,bb=0 100 500 500]{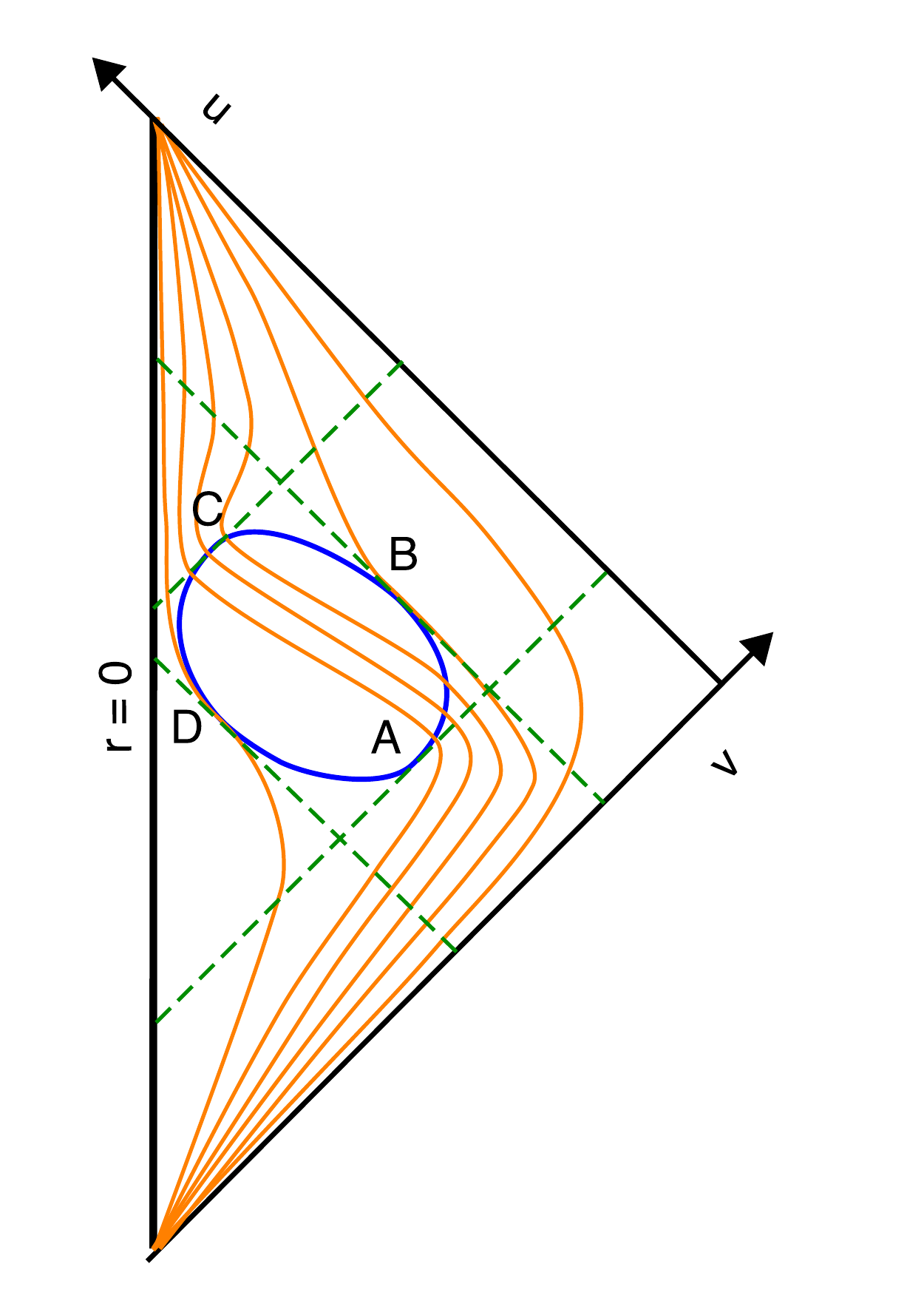}
\vskip2em
\caption{\small
The schematic Penrose diagram with an anti-trapping horizon.
\\
The interpretation of this diagram is analogous to
Fig.\ref{trapping-horizon-1}.
%The anti-trapping horizon is shown as a solid curve (in blue).
%The extremal points on the anti-trapping horizon
%with the largest/smallest values of the coordinate $u$ or $v$
%are marked as $A$, $B$, $C$ and $D$.
%The curves connecting the top and bottom vertices of the diagram
%are  constant-$r$ curves (in orange),
%and the dashed lines (in green) are constant-$u$ and constant-$v$ lines
%tangent to the trapping horizon.
}
\label{anti-trapping-horizon}
\vskip1em
\end{figure}

Let us first consider what would happen if 
there is actually an anti-trapping horizon.
All we need to do is to carry out a time-reversal transformation,
that is,
to interchange $u$ and $v$
in our discussions above.
Flipping the Penrose diagram Fig.\ref{trapping-horizon-1},
we find that for the anti-trapping horizon 
in Fig.\ref{anti-trapping-horizon}.

The location of the anti-trapping horizon 
can be specified as $u = u_0(v)$ 
for a given function $u_0(v)$.
Along the anti-trapping horizon, 
we must have
\begin{align}
\del_u r = 0,
\label{dur=0}
\end{align}
analogous to eq.\eqref{delvr=0}.
In particular,
for the segment $BC$ in Fig.\ref{anti-trapping-horizon},
generically we have
\begin{align}
\del^2_u r < 0,
\qquad
\del_v\del_u r < 0,
\label{du2r<0}
\end{align}
analogous to eq.\eqref{BC}.

The expansions \eqref{C-exp}, \eqref{r-exp} are changed to
\begin{align}
C(u, v) &=
\bar{C}_0(v) + \bar{C}_1(v)(u-u_0(v)) + \frac{1}{2}\bar{C}_2(v)(u-u_0(v))^2
%+ \frac{1}{6}\bar{C}_3(v)(u-u_0(v))^3 
+ \cdots,
\label{C-exp-1}
\\
r(u, v) &=
\bar{a}(v) + \frac{1}{2}\bar{r}_2(v)(u-u_0(v))^2
%+ \frac{1}{6}\bar{r}_3(v)(u-u_0(v))^3
%+ \frac{1}{24}\bar{r}_4(v)(u-u_0(v))^4 
+ \cdots.
\label{r-exp-1}
\end{align}

%Using the Einstein equations,
%one finds
%\begin{align}
%\bar{r}_2(v) &=
% \frac{1}{2}\kappa\bar{a}(v)\bar{T}^{(0)}_{uu}(v),
%\label{r2-1}
%\\
%u'_0(v) &=
%\frac{\bar{C}_0(v) - 2\kappa\bar{a}^2(v)\bar{T}^{(0)}_{uv}(v)}
%{- 2\kappa\bar{a}^2(v)\bar{T}^{(0)}_{uu}(v)},
%\label{primeu0-1}
%\end{align}
%where we have expanded the energy-momentum tensor as
%\be
%T_{\mu\nu}(u, v) = \bar{T}^{(0)}_{\mu\nu}(v) + \bar{T}^{(1)}_{\mu\nu}(v)(u-u_0(v))
%+ \frac{1}{2}\bar{T}^{(2)}_{\mu\nu}(v)(u-u_0(v))^2 + \cdots.
%\ee

%Due to eqs.\eqref{du2r<0} \eqref{r2-1},
Following similar calculations we did above
for the trapping horizon,
we find that
\begin{align}
\bar{r}_2(v) < 0
%\label{barr2<0}
\qquad
u'_0 < 0,
\label{u0prime<0}
\end{align}
implying that
\begin{align}
\bar{T}^{(0)}_{uu}(v) > 0,
%\label{barTuu>0}
\qquad
R_{\th\th}  > 1,
%\bar{C}_0(v) < 2\kappa\bar{a}^2(v)\bar{T}^{(0)}_{uv}(v)
\label{barC<barTuv}
\end{align}
where $\bar{T}^{(0)}_{uu}(v) \equiv T_{uu}(u_0(v), v)$,
as necessary conditions for the existence
of the segment $BC$ on the anti-trapping horizon.
Due to eq.\eqref{u0prime<0},
this segment of the anti-trapping horizon is space-like,
reflecting the fact that the inside
of the anti-trapping horizon is an anti-trapped region.

%%%PM04/10
As the condition \eqref{Rthth<<1} is necessarily violated,
we can no longer use these conditions to estimate
the energy-momentum tensor around
the anti-trapping horizon.
As the segment $BC$ of the anti-trapping horizon is space-like,
in fact, shrinking faster than light,
it is impossible for collapsing matter to
fall into the trapped region from here.

%%% End of hide

% example of figure and figure in the margin 
% can be found at the end of the file.

%\end{CJK} 

\vskip .8cm
\baselineskip 22pt
\begin{thebibliography}{99}
\itemsep 0pt



%\cite{Davies:1976ei}
\bibitem{Davies:1976ei} 
  P.~C.~W.~Davies, S.~A.~Fulling and W.~G.~Unruh,
  ``Energy-momentum Tensor Near an Evaporating Black Hole,''
  Phys.\ Rev.\ D {\bf 13}, 2720 (1976).
  doi:10.1103/PhysRevD.13.2720
  %%CITATION = doi:10.1103/PhysRevD.13.2720;%%
  %210 citations counted in INSPIRE as of 21 May 2016

%%%% WORMHOLE %%%%
%\cite{Parentani:1994ij}
\bibitem{Parentani:1994ij} 
  R.~Parentani and T.~Piran,
  ``The Internal geometry of an evaporating black hole,''
  Phys.\ Rev.\ Lett.\  {\bf 73}, 2805 (1994)
  doi:10.1103/PhysRevLett.73.2805
  [hep-th/9405007].
  %%CITATION = doi:10.1103/PhysRevLett.73.2805;%%
  %55 citations counted in INSPIRE as of 14 Dec 2016

%\cite{Ho:2018jkm}
\bibitem{Ho:2018jkm} 
  P.~M.~Ho and Y.~Matsuo,
  ``On the Near-Horizon Geometry of an Evaporating Black Hole,''
  JHEP {\bf 1807}, 047 (2018)
  doi:10.1007/JHEP07(2018)047
  [arXiv:1804.04821 [hep-th]].
  %%CITATION = doi:10.1007/JHEP07(2018)047;%%
  %2 citations counted in INSPIRE as of 02 Apr 2019

%\cite{Ho:2019cfw}
\bibitem{Ho:2019cfw} 
  P.~M.~Ho, Y.~Matsuo and S.~J.~Yang,
  ``Vacuum Energy at Apparent Horizon in Conventional Model of Black Holes,''
  arXiv:1904.01322 [hep-th].
  %%CITATION = ARXIV:1904.01322;%%

%\cite{Frolov:1981mz}
\bibitem{Frolov:1981mz} 
  V.~P.~Frolov and G.~A.~Vilkovisky,
  ``Spherically Symmetric Collapse in Quantum Gravity,''
  Phys.\ Lett.\ B {\bf 106}, 307 (1981).
  %%CITATION = PHLTA,B106,307;%%
  %91 citations counted in INSPIRE as of 09 Oct 2015

%\cite{Roman:1983zza}
\bibitem{Roman:1983zza} 
  T.~A.~Roman and P.~G.~Bergmann,
  ``Stellar collapse without singularities?,''
  Phys.\ Rev.\ D {\bf 28}, 1265 (1983).
  doi:10.1103/PhysRevD.28.1265
  %%CITATION = doi:10.1103/PhysRevD.28.1265;%%
  %59 citations counted in INSPIRE as of 14 May 2019

%\cite{Hayward:2005gi}
\bibitem{Hayward:2005gi} 
  S.~A.~Hayward,
  ``Formation and evaporation of regular black holes,''
  Phys.\ Rev.\ Lett.\  {\bf 96}, 031103 (2006)
  [gr-qc/0506126].
  %%CITATION = GR-QC/0506126;%%
  %152 citations counted in INSPIRE as of 09 Oct 2015

%%%% ORDER 1 CORRECTION AT HORIZON %%%%
%\cite{Mathur:2009hf}
\bibitem{Mathur:2009hf} 
  S.~D.~Mathur,
  ``The Information paradox: A Pedagogical introduction,''
  Class.\ Quant.\ Grav.\  {\bf 26}, 224001 (2009)
  [arXiv:0909.1038 [hep-th]].
  %%CITATION = ARXIV:0909.1038;%%
  %183 citations counted in INSPIRE as of 24 Apr 2015

%\cite{Marolf:2017jkr}
\bibitem{Marolf:2017jkr} 
  D.~Marolf,
  ``The Black Hole information problem: past, present, and future,''
  Rept.\ Prog.\ Phys.\  {\bf 80}, no. 9, 092001 (2017)
  doi:10.1088/1361-6633/aa77cc
  [arXiv:1703.02143 [gr-qc]].
  %%CITATION = doi:10.1088/1361-6633/aa77cc;%%
  %51 citations counted in INSPIRE as of 01 May 2019

%\cite{Kawai:2013mda}
\bibitem{Kawai:2013mda} 
  H.~Kawai, Y.~Matsuo and Y.~Yokokura,
  ``A Self-consistent Model of the Black Hole Evaporation,''
  Int.\ J.\ Mod.\ Phys.\ A {\bf 28}, 1350050 (2013)
  [arXiv:1302.4733 [hep-th]].
  %%CITATION = ARXIV:1302.4733;%%
  %8 citations counted in INSPIRE as of 31 Mar 2015
  
%\cite{Kawai:2014afa}
\bibitem{Kawai:2014afa} 
  H.~Kawai and Y.~Yokokura,
  ``Phenomenological Description of the Interior of the Schwarzschild Black Hole,''
  Int.\ J.\ Mod.\ Phys.\ A {\bf 30}, 1550091 (2015)
  doi:10.1142/S0217751X15500918
  [arXiv:1409.5784 [hep-th]].
  %%CITATION = doi:10.1142/S0217751X15500918;%%
  %12 citations counted in INSPIRE as of 12 Nov 2016

%\cite{Ho:2015fja}
\bibitem{Ho:2015fja} 
  P.~M.~Ho,
  ``Comment on Self-Consistent Model of Black Hole Formation and Evaporation,''
  JHEP {\bf 1508}, 096 (2015)
  doi:10.1007/JHEP08(2015)096
  [arXiv:1505.02468 [hep-th]].
  %%CITATION = doi:10.1007/JHEP08(2015)096;%%
  %4 citations counted in INSPIRE as of 12 Nov 2016

%\cite{Kawai:2015uya}
\bibitem{Kawai:2015uya} 
  H.~Kawai and Y.~Yokokura,
  ``Interior of Black Holes and Information Recovery,''
  Phys.\ Rev.\ D {\bf 93}, no. 4, 044011 (2016)
  doi:10.1103/PhysRevD.93.044011
  [arXiv:1509.08472 [hep-th]].
  %%CITATION = doi:10.1103/PhysRevD.93.044011;%%
  %11 citations counted in INSPIRE as of 12 Nov 2016

%\cite{Ho:2015vga}
\bibitem{Ho:2015vga} 
  P.~M.~Ho,
  ``The Absence of Horizon in Black-Hole Formation,''
  Nucl.\ Phys.\ B {\bf 909}, 394 (2016)
  doi:10.1016/j.nuclphysb.2016.05.016
  [arXiv:1510.07157 [hep-th]].
  %%CITATION = doi:10.1016/j.nuclphysb.2016.05.016;%%
  %6 citations counted in INSPIRE as of 12 Nov 2016

%\cite{Ho:2016acf}
\bibitem{Ho:2016acf} 
  P.~M.~Ho,
  ``Asymptotic Black Holes,''
  Class.\ Quant.\ Grav.\  {\bf 34}, no. 8, 085006 (2017)
  doi:10.1088/1361-6382/aa641e
  [arXiv:1609.05775 [hep-th]].
  %%CITATION = doi:10.1088/1361-6382/aa641e;%%
  %4 citations counted in INSPIRE as of 20 Apr 2017

%\cite{Kawai:2017txu}
\bibitem{Kawai:2017txu} 
  H.~Kawai and Y.~Yokokura,
  ``A Model of Black Hole Evaporation and 4D Weyl Anomaly,''
  arXiv:1701.03455 [hep-th].
  %%CITATION = ARXIV:1701.03455;%%
  %3 citations counted in INSPIRE as of 20 Apr 2017

%\cite{Stephens:1993an}
\bibitem{Stephens:1993an} 
  C.~R.~Stephens, G.~'t Hooft and B.~F.~Whiting,
  ``Black hole evaporation without information loss,''
  Class.\ Quant.\ Grav.\  {\bf 11}, 621 (1994)
  doi:10.1088/0264-9381/11/3/014
  [gr-qc/9310006].
  %%CITATION = doi:10.1088/0264-9381/11/3/014;%%
  %396 citations counted in INSPIRE as of 02 Apr 2019

%%%% FUZZBALL %%%%
\bibitem{FuzzBall}
%\cite{Lunin:2001jy}
%\bibitem{Lunin:2001jy} 
  O.~Lunin and S.~D.~Mathur,
  ``AdS / CFT duality and the black hole information paradox,''
  Nucl.\ Phys.\ B {\bf 623}, 342 (2002)
  [hep-th/0109154].
  %%CITATION = HEP-TH/0109154;%%
  %235 citations counted in INSPIRE as of 24 Apr 2015
%\cite{Lunin:2002qf}
%\bibitem{Lunin:2002qf} 
  O.~Lunin and S.~D.~Mathur,
  ``Statistical interpretation of Bekenstein entropy for systems with a stretched horizon,''
  Phys.\ Rev.\ Lett.\  {\bf 88}, 211303 (2002)
  [hep-th/0202072].
  %%CITATION = HEP-TH/0202072;%%
  %125 citations counted in INSPIRE as of 24 Apr 2015

%%%% FIREWALL %%%%
\bibitem{firewall}
%\cite{Almheiri:2012rt}
%\bibitem{Almheiri:2012rt} 
  A.~Almheiri, D.~Marolf, J.~Polchinski and J.~Sully,
  ``Black Holes: Complementarity or Firewalls?,''
  JHEP {\bf 1302}, 062 (2013)
  [arXiv:1207.3123 [hep-th]];
  %%CITATION = ARXIV:1207.3123;%%
  %314 citations counted in INSPIRE as of 31 Mar 2015
%\bibitem{Braunstein}
S. L. Braunstein, 
``Black hole entropy as entropy of entanglement, 
  or it's curtains for the equivalence principle,''
[arXiv:0907.1190v1 [quant-ph]] 
published as 
%\cite{Braunstein:2009my}
%\bibitem{Braunstein:2009my} 
  S.~L.~Braunstein, S.~Pirandola and K.~Życzkowski,
  ``Better Late than Never: Information Retrieval from Black Holes,''
  Phys.\ Rev.\ Lett.\  {\bf 110}, no. 10, 101301 (2013),
  %[arXiv:0907.1190 [quant-ph]].
  %%CITATION = ARXIV:0907.1190;%%
  %170 citations counted in INSPIRE as of 13 May 2015
  for a similar prediction from different assumptions.

\cite{Hooft:2018gtw}
\bibitem{Hooft:2018gtw} 
 G.~'t Hooft,
 ``What happens in a black hole when a particle meets its antipode,''
arXiv:1804.05744 [gr-qc].
  %%CITATION = ARXIV:1804.05744;%%
  %2 citations counted in INSPIRE as of 02 Apr 2019
















\end{thebibliography}
\end{document}